\documentclass{eas}
\usepackage{graphicx}
\begin{document}

\runningtitle{Torres: Observational constraints from binary stars \dots}
\title{Observational Constraints From Binary Stars on Stellar Evolution Models} 
\author{Guillermo Torres}\address{Harvard-Smithsonian Center for
Astrophysics, 60 Garden St., Cambridge MA 02138, USA}
\begin{abstract}
Accurate determinations of masses and radii in binary stars, along
with estimates of the effective temperatures, metallicities, and other
properties, have long been used to test models of stellar evolution.
As might be expected, observational constraints are plentiful for
main-sequence stars, although some problems with theory remain even in
this regime. Models in other areas of the H-R diagram are considerably
less well constrained, or not constrained at all.  I summarize the
status of the field, and provide examples of how accurate measurements
can supply stringent tests of stellar theory, including aspects such
as the treatment of convection. I call attention to the apparent
failure of current models to match the properties of stars with masses
of 1.1--1.7\,$M_{\odot}$ that are near the point of central hydrogen
exhaustion, possibly connected with the simplified treatment of
convective core overshooting.
\end{abstract}
\maketitle
\section{Introduction}

Stellar evolution theory represents the backbone of much of modern
Astrophysics. For decades astronomers have worked to gather
observations of many different kinds to constrain and test various
physical ingredients of the models, and to calibrate a number of free
parameters. These include the helium abundance, mass loss rates, and
convective quantities such as the mixing length parameter
($\alpha_{\rm ML}$) and the amount of overshooting from the convective
core ($\alpha_{\rm ov}$). In the last decade or two, accurate
observations have revealed several shortcomings in our knowledge of
stellar physics. One example is the difficulty in reproducing the
radii and effective temperatures of late-type stars, which are larger
and cooler than predicted by current standard models (see, e.g.,
Torres \cite{Torres:13a}, and also the contribution by Greg Feiden in
these Proceedings). Below I will describe another problem that is
possibly related to the treatment of convection.

Perhaps one of the best known ways of constraining stellar evolution
theory is by means of color-magnitude diagrams (CMDs) of star
clusters, which have been compared extensively with model isochrones
to infer other interesting properties of the clusters such as age,
distance, or chemical composition. Comparisons like these are powerful
probes of stellar physics, but are not without their
difficulties. Contamination of the CMDs by field stars, or unresolved
binaries, can complicate or bias the analysis, as can stellar
variability and reddening.  An additional source of uncertainty is the
color-temperature transformations used to convert models from the
theoretical plane to the observational plane. The technique of
asteroseismology provides very different but highly complementary
observational constraints, through the measurement of oscillation
frequencies that give us access to properties of the stellar
interiors.  These are challenging measurements, however, which
typically require high-precision, continuous, and long-term
observations, and are generally best done in bright stars with
luminosities similar to the Sun or higher. A third, important way to
test models that is again complementary to the previous two is through
the observation of detached binary systems, which enable the
model-independent measurement of fundamental stellar properties such
as the mass and radius, and also effective temperature, luminosity,
etc. While simple in principle, this technique requires special
configurations and is not always easy for all types of stars.

In this paper I will focus on how binary stars can help to test
aspects of stellar evolution theory. It is useful to begin by
reviewing the status of fundamental mass and radius determinations in
eclipsing binaries, as recorded in the handful of ``critical'' reviews
that have appeared in the literature. These are compilations that pay
special attention not only to the formal precision of the
measurements, but also to the quality of the data and the analysis,
particularly regarding systematic errors. The first critical review by
Popper (\cite{Popper:67}) listed only two systems with mass
determinations (but no radii) having relative errors under 3\%. A
subsequent compilation by Popper (\cite{Popper:80}) increased this to
7 systems with masses and radii good to the same accuracy. Andersen
(\cite{Andersen:91}) brought the total to 45 systems, and Torres
\etal\ (\cite{Torres:10}) more than doubled it, to 95 systems. The
masses and radii for some of these systems, along with other measured
properties, allow for very stringent tests of models, as illustrated
in the latter two references. Here I will concentrate on the
phenomenon of overshooting from the convective core.

\section{Convective core overshooting: how binaries can help}

Overshooting can be understood as mixing beyond the boundary of the
convective core as given by the classical Schwarzschild
(\cite{Schwarzschild:06}) criterion: rising convective elements
``overshoot'' into the radiative zone. There are a number of important
consequences of overshooting that affect the later stages of
evolution. Enhanced mixing prolongs core hydrogen burning by feeding
more H-rich material into the core. This changes the ages predicted by
models. Access to a larger hydrogen reservoir during the H-burning
phase enhances the mass of the helium core left behind, and this
alters the global characteristics of the giant phases. In particular,
it shortens the shell H-burning phase, reduces the lifetime of the
core He-burning phase, and affects the luminosities in the giant
stages. The effect on the main sequence portion of the evolutionary
tracks in the H-R diagram is to extend the tracks toward cooler
temperatures and higher luminosities, as illustrated, e.g., by
Schr\"oder \etal\ (\cite{Schroder:97}). All giant phases occur at
higher luminosities than they would without overshooting. For massive
stars even the pre-main sequence (PMS) phases are affected, as shown
by Marques \etal\ (\cite{Marques:06}): the evolutionary track for a
4\,$M_{\odot}$ PMS star develops an extra loop near the zero-age main
sequence that is completely absent if overshooting is not considered.

Even though there has been considerable progress in understanding
turbulent convection, the sizes of convective cores in stars still
cannot be predicted from first principles (VandenBerg \etal\
\cite{VandenBerg:06}). The most common approach in stellar evolution
models is to parametrize the effect of overshooting in terms of a
single variable representing the length of overshooting as a function
of $H_p$, the local pressure scale height: $l_{\rm ov} = \alpha_{\rm
ov} H_p$. This formulation is easy to implement, but the overshooting
parameter $\alpha_{\rm ov}$ must be calibrated using observations.
This can be done in several ways. One is to use CMDs of clusters. In
this approach one tries to match the detailed shape and extent of the
``blue hook'' region by adjusting $\alpha_{\rm ov}$, as illustrated by
Demarque \etal\ (\cite{Demarque:04}). This method works quite well,
though it is somewhat vulnerable to the uncertainties mentioned
earlier. Another technique takes advantage of accurate measurements of
the masses, radii, and temperatures of eclipsing binaries that are
near the end of their main-sequence life. It is based on the premise
that the likelihood of finding a random field star in the region of
the H-R diagram corresponding to the shell H-burning phase is small,
because evolution across this so-called ``Hertzprung gap'' is very
rapid.  Therefore, if a star appears to be slightly beyond the point
of hydrogen exhaustion as marked by an evolutionary track, it is
usually possible to increase the amount of overshooting in the models
so that the track ``reaches out'' to the star, bringing it onto a
location still on the main sequence that is \apriori\ much more
likely. An example of this procedure is seen in the study of the
eclipsing binary GX\,Gem by Lacy \etal\ ({\cite{Lacy:08}), shown in
Figure~\ref{fig:GXGem}.  Caveats are that this procedure can be
sensitive to errors in the measured temperatures, and that there is a
certain amount of degeneracy with metallicity, if it is not known
observationally for the system.

\begin{figure}[t!]
\vspace{0.0in}\vbox{
\begin{tabular}{cc}
\begin{minipage}[h]{3.5in}
\hspace{-0.08in}\includegraphics[width=7cm]{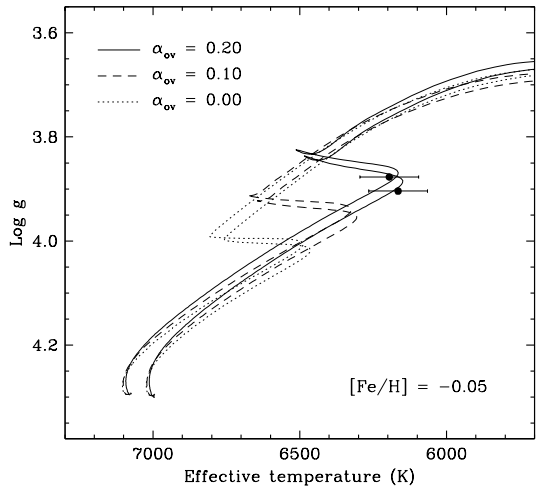}
\end{minipage} &
\hspace{-0.9in}
\begin{minipage}[h]{1.9in}
\caption{Illustration adapted from Figure~8 by Lacy \etal\
(\cite{Lacy:08}) of the constraint on the overshooting parameter
provided by the detached eclipsing binary GX\,Gem, which is composed
of two evolved F stars with measured masses of 1.49 and
1.47\,$M_{\odot}$.  Evolutionary tracks are shown here for these
masses from the models by Claret (\cite{Claret:04}), for a metal
abundance ${\rm [Fe/H]} = -0.05$, and suggest that the best fit to the
observations is obtained with an overshooting parameter near
$\alpha_{\rm ov} = 0.2$. \label{fig:GXGem}}
\end{minipage}
\end{tabular}
}
\end{figure}

Typical values of $\alpha_{\rm ov}$ are in the range 0.1--0.2. The
treatment of overshooting in the transition region where stars begin
to develop convective cores (approximately the mass range
1.1--1.7\,$M_{\odot}$) is particularly difficult. The ways in which
different models ramp up the overshooting from zero to some maximum
value varies from model to model, but they are all rather arbitrary
and therefore a source of concern. For example, in the Yonsei-Yale
models (Yi \etal\ \cite{Yi:01}) the overshooting parameter is
increased in steps of 0.05 starting at some mass value $M^{\rm
conv}_{\rm crit}$ at which stars develop cores, which is
metallicity-dependent. It is then held constant at the value
$\alpha_{\rm ov} = 0.2$ for masses above $M^{\rm conv}_{\rm crit} +
0.2\,M_{\odot}$ (Demarque \etal\ \cite{Demarque:04}). The
Victoria-Regina models (VandenBerg \etal\ \cite{VandenBerg:06}) use a
somewhat different prescription for overshooting that is equivalent to
the single-parameter formalism used in the Yonsei-Yale models, and
ramps up the strength of the overshooting in a smoother but different
way, which also depends on metallicity.

A persistent question has been whether and exactly how $\alpha_{\rm
ov}$ depends on stellar mass. Schr\"oder \etal\ (\cite{Schroder:97})
used accurate measurements for binary systems containing giant or
supergiant primaries to estimate the degree of overshooting in the
same way described above over the mass range 2--8\,$M_{\odot}$, and
concluded that $\alpha_{\rm ov}$ increases from about 0.2 to 0.3 over
this interval. A similar study by Ribas \etal\ (\cite{Ribas:00}) used
eight main-sequence eclipsing binaries with components ranging from 2
to 12\,$M_{\odot}$, and also found a systematic increase in
$\alpha_{\rm ov}$, consistent with the previous results. However, the
more recent study by Claret (\cite{Claret:07}) based on a larger
number of main-sequence eclipsing binaries (13) with masses between 2
and 30\,$M_{\odot}$ found a much shallower dependence of $\alpha_{\rm
ov}$ on mass that is also considerably more uncertain, so the question
of the mass dependence of overshooting, if any, remains.

\section{Constraints on overshooting from the eclipsing binary AQ\,Ser:
indications of a new problem with stellar evolution models}

Another illustration of how eclipsing binaries can help to calibrate
models is provided by the F-star system AQ\,Ser, which has component
masses of about 1.42 and 1.35\,$M_{\odot}$, and relative errors
smaller than 2\% in both the masses and radii (Torres \etal\
\cite{Torres:13b}). This is a highly evolved system presumably at the
very end of its main-sequence phase, which makes it uniquely sensitive
for testing convective core overshooting. A similar exercise as
described earlier for GX\,Gem constrains $\alpha_{\rm ov}$ to be
between 0.2 and 0.3 (see Figure~\ref{fig:AQSer}, left panels), using
the Granada stellar evolution models by Claret
(\cite{Claret:04}). However, the models are unable to match the well
measured temperature \emph{difference} between the components ($\Delta
T_{\rm eff} = 90 \pm 20$\,K), even in sign: they predict the primary
star to be hotter, while the observations indicate the reverse.

\begin{figure}
\vspace{0.0in}\vbox{
\begin{tabular}{cc}
\begin{minipage}[h]{3.5in}
\hspace{-0.02in}\includegraphics[width=5.1cm]{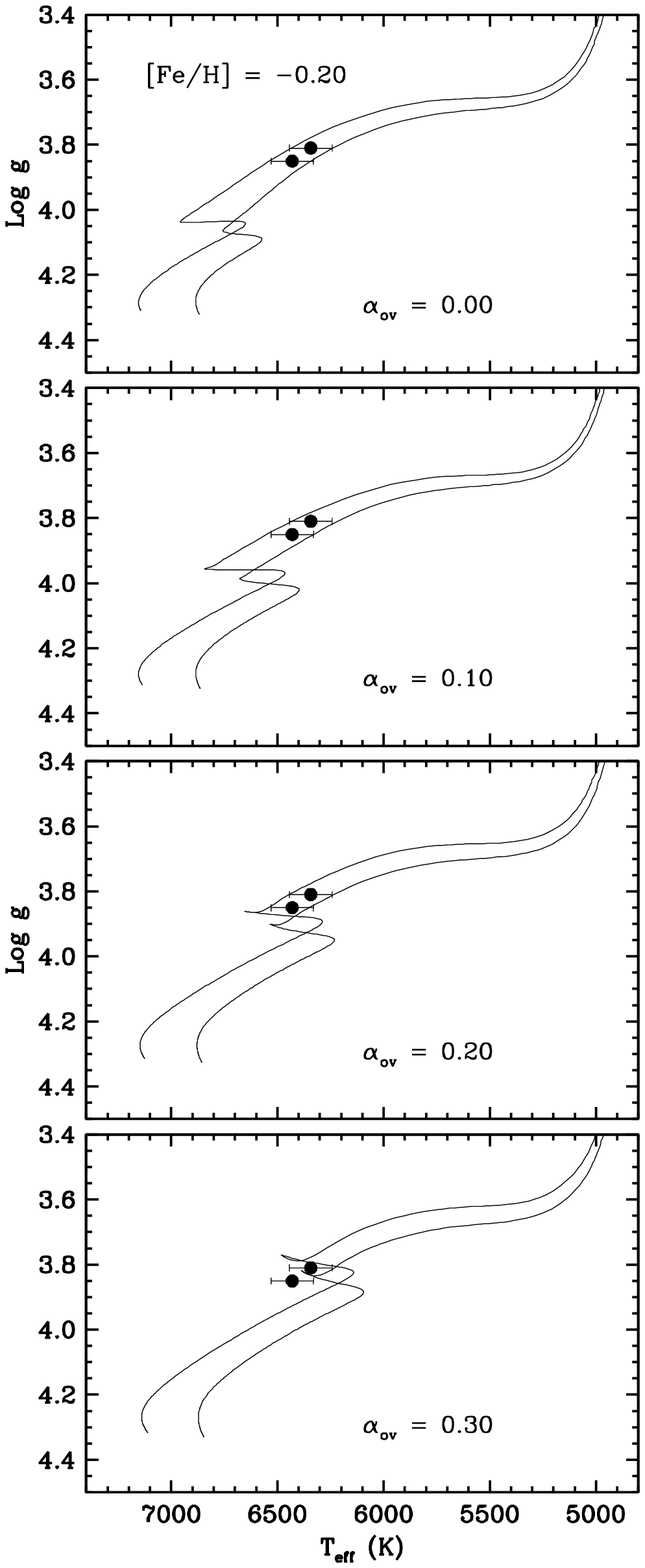}
\end{minipage} &
\hspace{-1.3in}
\begin{minipage}[h]{1.9in}
\hspace{-0.27in}\includegraphics[width=6.82cm]{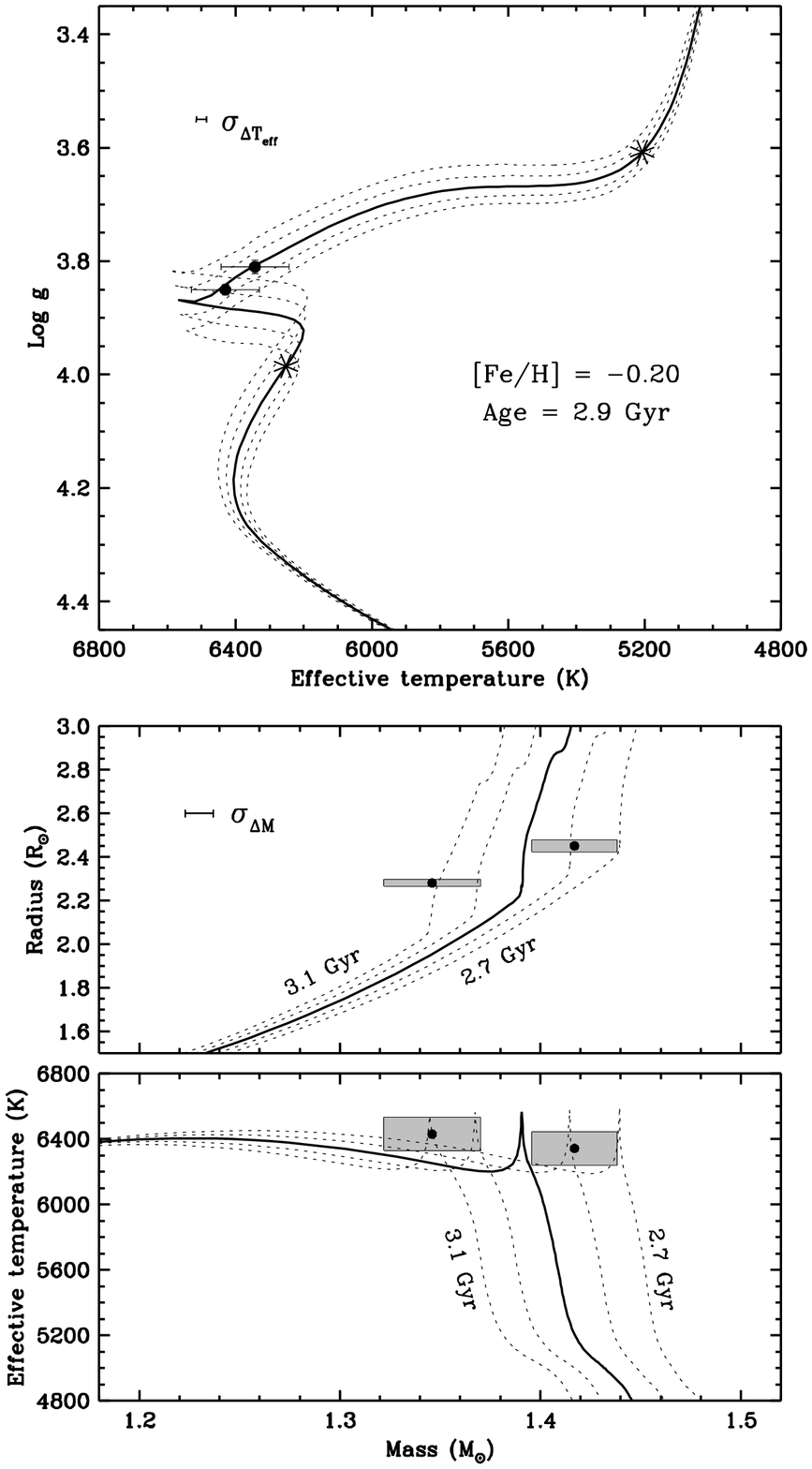}
\end{minipage}
\end{tabular}
}
\caption{Observations of the eclipsing binary AQ\,Ser compared with
stellar evolution models (figure adapted from Torres \etal\
\cite{Torres:13b}). \emph{Left:} Mass tracks by Claret
(\cite{Claret:04}) for the primary and secondary components, for a
range of overshooting parameters. The best match is near $\alpha_{\rm
ov} = 0.3$ for ${\rm [Fe/H]} = -0.20$. \emph{Top right:} Isochrones
from the models by VandenBerg \etal\ (\cite{VandenBerg:06}),
suggesting a satisfactory match for the same metallicity and an age of
2.9 Gyr. However, the predicted locations of the stars for the
measured masses, indicated by the asterisks on the best-fit isochrone
(solid line), are inconsistent with the observations. \emph{Bottom
right:} Radius and effective temperature as a function of mass, along
with calculations from VandenBerg \etal\ (\cite{VandenBerg:06}). The
models predict a younger age for the primary star. \label{fig:AQSer}}
\end{figure}

This discrepancy manifests itself in other ways, independently of the
model considered. For example, the top-right panel of
Figure~\ref{fig:AQSer} shows the best-fit isochrone from the
Victoria-Regina series. In these models the strength of overshooting
is fixed and cannot be changed by the user. The models seemingly match
the observations very well for an age of 2.9 Gyr at a metallicity of
${\rm [Fe/H]} = -0.20$, and suggest the stars are slightly beyond the
point of hydrogen exhaustion. However, the predicted location of the
stars on this isochrone from their nominally measured masses, which is
indicated with the asterisks, is very far from the actual locations
marked by the filled circles and error bars. In other words, the
models would predict a mass ratio much closer to unity ($q \equiv
M_2/M_1 = 1.00083$) than that measured spectroscopically ($q = 1.054
\pm 0.011$). The difference is highly significant, at nearly the
5-$\sigma$ level. Yet another way to visualize the disagreement is
presented in the lower-right panels of Figure~\ref{fig:AQSer},
particularly in the diagram of radius versus mass. It is seen that the
models predict the more massive primary star to be younger than the
secondary, by about 0.3 Gyr (10\%). Similar discrepancies in the same
direction are obtained with the Yonsei-Yale and Granada models (0.45
Gyr and 0.5 Gyr, respectively). Experiments with several different
trial values of [Fe/H] in which $\alpha_{\rm ov}$ and also
$\alpha_{\rm ML}$ are allowed to vary independently for each star do
not improve the age agreement, pointing to a fundamental problem with
the models.

As it turns out, similar discrepancies have been reported by Clausen
\etal\ (\cite{Clausen:10}) for a handful similarly evolved F stars:
GX\,Gem, V442\,Cyg, BW\,Aqr, and BK\,Peg. All yield predicted ages for
the primary stars that are younger than the secondaries, no matter
which model is used. Two other systems showing the same anomaly have
been identified more recently, although the problem was not noted in
the original publications: CO\,And (Lacy \etal\ {\cite{Lacy:10}) and
BF\,Dra (Lacy \etal\ {\cite{Lacy:12}).

\begin{figure}
\vspace{1.2in}
\begin{center}
\includegraphics[width=7.2cm]{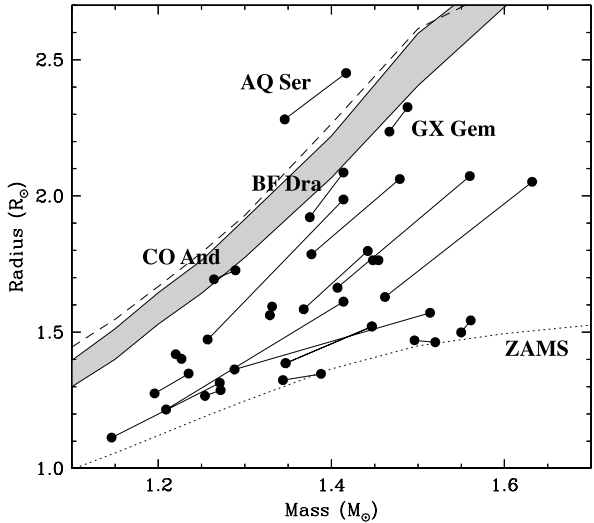}
\end{center}
\vspace{-0.25in}
\caption{Masses and radii for all eclipsing binaries with accurately
known parameters from Torres \etal\ (\cite{Torres:10}), supplemented
with measurements for CO\,And (Lacy \etal\ \cite{Lacy:10}), BF\,Dra
(Lacy \etal\ \cite{Lacy:12}), and AQ\,Ser (Torres \etal\
\cite{Torres:13b}). The primary and secondary stars in each system are
connected with a line. The shaded area corresponds to the blue hook
region for solar metallicity, according to the Yonsei-Yale models. The
dashed line above it represents the upper envelope of this region for
a metallicity of ${\rm Fe/H]} = -0.20$ that is close to that of
AQ\,Ser, and the dotted line at the bottom corresponds to the zero-age
main sequence (ZAMS). \label{fig:evolved}}
\end{figure}

A common property of these binary systems is that the components all
have masses in the range of 1.1--1.7\,$M_{\odot}$, which is precisely
where models ramp up the importance of overshooting, and they are all
considerably evolved (i.e., near the end of the main-sequence
phase). This is also the mass range in which stars transition from
having their energy production dominated by the p-p cycle to the CNO
cycle. Figure~\ref{fig:evolved} displays all binary systems with well
measured properties that have masses in the range indicated above.
The shaded area represents the region of the blue hook, according to
the Yonsei-Yale models, and an increase in $\alpha_{\rm ov}$ would
shift this region upwards.  AQ\,Ser is seen to be the most evolved
system in this regime, and is also the one showing the most pronounced
discrepancy with theory.

Given that convective core overshooting has a direct impact on
evolution timescales, especially for main-sequence stars in the more
advanced stages, it is natural to suspect that the simplified
treatment of this phenomenon in current stellar models has something
to do with their failure to reproduce the observed properties of
well-measured eclipsing binaries at a single age. However, from the
experiments with AQ\,Ser described above, the explanation does not
appear to be a simple difference in $\alpha_{\rm ov}$ for the two
components, and may be more complex. At the very least, it may involve
a dependence of overshooting on the state of evolution, in addition to
mass and metallicity.

This problem has not received much attention in the literature beyond
the work of Clausen \etal\ (\cite{Clausen:10}), and is a good example
of the usefulness of accurate measurements of eclipsing binaries for
testing our knowledge of stellar evolution.

\acknowledgements

This work was supported in part by grant AST-1007992 from the US
National Science Foundation.


\end{document}